\documentclass{article}
\usepackage{arxiv}
\usepackage[utf8]{inputenc}
\usepackage[T1]{fontenc}
\usepackage{hyperref}
\usepackage{url}
\usepackage{booktabs}
\usepackage{amsfonts}
\usepackage{microtype}
\usepackage{graphicx}
\usepackage{amsmath,amssymb}
\usepackage{natbib}
\usepackage{tabularx}
\usepackage{array}

\title{TypedInk: Knowing When Not to Draw}
\author{Weiwei Yang\\
Solus\textsuperscript{1}\\
\texttt{wxs123x@gmail.com}}
\renewcommand{\shorttitle}{TypedInk}

\hypersetup{
  pdftitle={TypedInk: Knowing When Not to Draw},
  pdfauthor={Weiwei Yang},
  pdfsubject={Scientific illustration from incomplete evidence; evidence-typed, integrity-gated figure generation},
  pdfkeywords={scientific illustration, incomplete evidence, typed rendering, academic diagrams, statistical plots}
}

\begin{document}
\maketitle
\footnotetext[1]{\url{https://solus.xin}}

\begin{abstract}
Can a scientific figure generator know when it does not have enough evidence to draw?
Complete-source illustrators map a finished methodology section to a schematic, or a table to plotting code, and are never asked this question because their input is assumed complete.
The research record is rarely that complete: an outline, a draft paragraph, a table, or a one-sentence intent, in any combination, and painting measured numbers with an image model when the table is missing is a hallucinated result.
TypedInk infers a figure type $\tau\in\{\mathrm{diagram},\mathrm{plot},\mathrm{plate}\}$ from a possibly incomplete source $S$ and dispatches to the renderer that can honor its evidence contract $\kappa(\tau)$, refusing dispatch when the contract cannot be met.
On the six gold plots in TypedFigBench (5 views, $n=30$), TypedInk's code path inks every gold table token when a table is observed ($12/12$), refuses the $12$ hidden-table views instead of inventing numbers, and marks $6$ intent-only views illustrative, with no execution failures.
An always-draw coder with no such gate inks $30/30$ charts, including all $18$ with no table, and recovers zero gold numbers on any of them; two vision-language raters (not humans; human replication is future work) still judge those always-draw charts submittable $17/36$ times, against TypedInk's refusal $0/36$ times.
A diagnostic field ablation of the type router (16 illustrations $\times$ 5 views, $n=80$) locates where this failure would originate if unchecked: an outline alone yields type accuracy $0.31$ and sends every gold plot into an integrity violation, and a chat-model planner narrows but does not close the gap (overall accuracy $0.85$, still integrity-failing on $25\%$ of outlines).
Diagram-label recall is $0.47$ on the full record and $0.19$ on an outline.
Gold plates are typed at dispatch (oracle integrity-fail $0.00$) and rendered as stills.
\end{abstract}


\section{Introduction}
\label{sec:intro}

Autonomous scientific agents can retrieve literature, propose experiments, and iterate on code~\citep{lu2024aiscientist,gottweis2025coscientist,langley1987discovery}.
Visual communication remains a bottleneck~\citep{zhu2026paperbanana}.
A discovery that cannot be drawn to the standards of an academic manuscript is only half communicated.

Recent work treats that bottleneck as text-to-figure generation under a \emph{complete} source.
PaperBanana~\citep{zhu2026paperbanana} writes $I=f(S,C)$ with $S$ a methodology section and $C$ a caption, then orchestrates retrieval, planning, style, rendering, and critique.
AutoFigure~\citep{zhu2026autofigure} maps a finished long document (avg.\ ${>}10$k tokens) to one conceptual schematic: a symbolic SVG/HTML layout, an aesthetic renderer, and an erase-and-correct text overlay.
FigureBench, their $3{,}300$-pair bench, retains conceptual illustrations and \emph{explicitly excludes data-driven charts}.
Even then, AutoFigure's win-rate against the human figure on research papers is $53\%$.
SciFig~\citep{huang2026scifig} maps long scientific text to optionally editable XML.
Code-based systems map tables to matplotlib~\citep{yang2024matplotagent,yang2025chartmimic}.
TikZ synthesizers address vector graphics from text or sketches~\citep{belouadi2023automatikz,belouadi2024detikzify}.
All of these lines assume a source that is already typed: a method section when the target is a diagram, a table when the target is a plot.
Table~\ref{tab:compare} places TypedInk against the three complete-source illustrators on that assumption.
Most of the table stays a protocol comparison, since AutoFigure structurally excludes data-driven charts and SciFig treats programming-based plots as an already-solved adjacent problem, so neither offers an untyped-plot baseline to run; PaperBanana does define a plot Visualizer, and its bottom row runs PaperBanana's own Visualizer prompt and execution code directly (Appendix~\ref{app:naive}).

That assumption does not hold in the research workflow.
Before a methodology section exists, a scientist holds fragments: an outline of claims that does not entail a topology, a paragraph of mechanism that does not entail a measured bar height, a table of LVEF values that does not entail a TEM plate.
Collapsing these fragments into one image-generation prompt produces two failures that aesthetics metrics miss: (i)~measured quantities painted by an image model, and (ii)~in-hand specimens rendered as cinematic stills or as charts.
Knowing when not to answer is a studied problem in machine learning under selective prediction and abstention~\citep{elyaniv2010selective,wen2024knowyourlimits}; we bring that same discipline to scientific illustration, where the thing withheld under insufficient evidence is not an answer but unsupported visual content.

\begin{figure}[t]
  \centering
  \includegraphics[width=\linewidth]{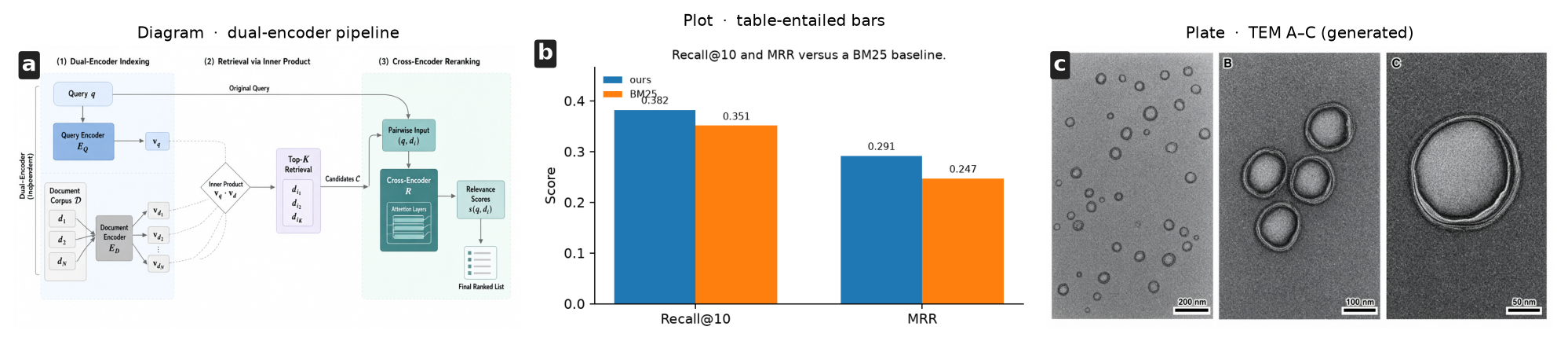}
  \caption{Three evidence contracts, rendered by TypedInk.
  (a)~Diagram: dual-encoder retrieval with a cross-encoder reranker.
  (b)~Plot: Recall@10 and MRR from the gold table; every labeled number is in $S_{\mathrm{table}}$.
  (c)~Plate: TEM panels A--C with scale bars.
  (b) is the code backend; (a) and (c) are the still template.}
  \label{fig_families}
\end{figure}

\begin{figure}[p]
  \centering
  \includegraphics[width=\linewidth,height=0.78\textheight,keepaspectratio]{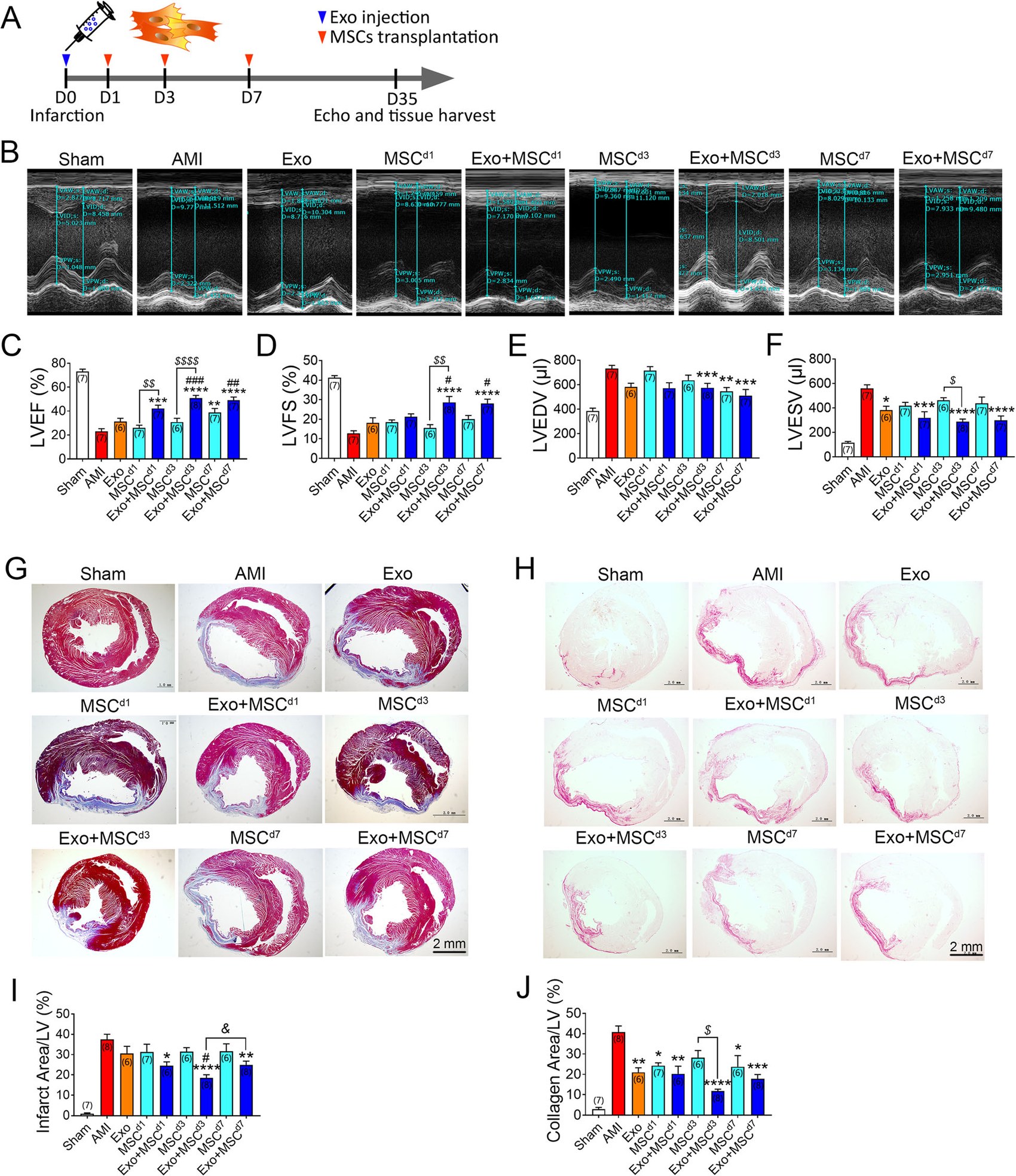}
  \caption{A published mixed-type figure, reproduced from Huang et al., \emph{Stem Cell Research \& Therapy} 10:300 (2019), Fig.~2~\citep{huang2019combinatorial}, under CC-BY~4.0.
  Panel~A is a protocol schematic (diagram); B, G, H are observational plates (echocardiogram, Masson trichrome, Sirius Red); C--F, I, J are measured plots (LVEF, volumes, infarct and collagen fractions).
  One caption, three evidence contracts.
  Extracted from the publisher PDF and resized for this layout; panel content unaltered.
  This is not a TypedInk output.}
  \label{fig_journal}
\end{figure}

We argue that scientific figures are \emph{typed artifacts with evidence contracts} (Figure~\ref{fig_families}).
Journals combine mixed types under one caption (Figure~\ref{fig_journal}).
TypedInk infers the type from whatever subset of the record is observed, refuses renderers that would break the contract, and composes the page (Figure~\ref{fig_compose}).

This paper makes four contributions.
\begin{enumerate}
\item \textbf{Task.} Illustration from incomplete evidence, $S=(S_{\mathrm{struct}},S_{\mathrm{prose}},S_{\mathrm{table}},S_{\mathrm{intent}})$, any field of which may be empty, together with a communicative intent $C$ (Section~\ref{sec:task}).
\item \textbf{Evidence contracts.} Three figure types $\tau\in\{\mathrm{diagram},\mathrm{plot},\mathrm{plate}\}$, each carrying a contract $\kappa(\tau)$ on what evidence licenses its content, and an integrity predicate distinct from type error (Section~\ref{sec:task}).
\item \textbf{Integrity-gated renderer.} TypedInk, $\hat\tau=\Pi(S,C)$, $I=R_{\hat\tau}(S,C)$: a compiler that refuses measured ink without a table, recovers gold numbers when the table is present, treats plates as a typed family at dispatch, and rolls $\hat\tau$ back to code whenever a gold plot would otherwise reach an image model (Section~\ref{sec:method}).
\item \textbf{A benchmark that measures hallucinated visual evidence}, not only type accuracy: gold-table recall and refuse-versus-always-draw on live generation ($n=30$), plus two-rater VL integrity judgments ($n=36$ hidden-table, $n=24$ measured, $n=12$ plate). The 16-illustration TypedFigBench field ablation ($n=80$) is reported separately, as a diagnostic of where the type router fails, not as an independent classification result (Section~\ref{sec:experiments}).
\end{enumerate}

\section{Task Formulation}
\label{sec:task}

Let $\mathcal{R}$ be the latent scientific record: claims, prose, measurements, and in-hand objects.
An agent observes a \emph{source}
\begin{equation}
S = (S_{\mathrm{struct}},\, S_{\mathrm{prose}},\, S_{\mathrm{table}},\, S_{\mathrm{intent}}),
\end{equation}
any component of which may be empty, together with a communicative intent $C$ (often $S_{\mathrm{intent}}$ itself).
This is PaperBanana's mapping~\citep{zhu2026paperbanana} with two changes: $S$ is a partial observation, and the output is typed.

\begin{figure}[t]
  \centering
  \includegraphics[width=\linewidth]{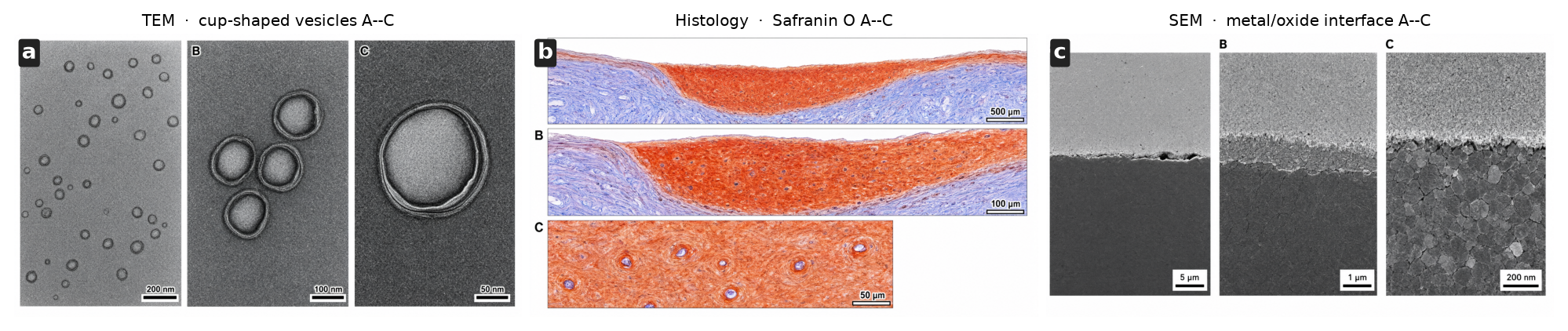}
  \caption{Observational plates under the still template (Qwen-Image-3.0-Pro).
  (a)~TEM of cup-shaped vesicles, panels A--C.
  (b)~Safranin O histology of repaired cartilage, A--C.
  (c)~Cross-sectional SEM of an annealed metal/oxide interface, A--C.
  Generated rasters, not photographs from the authors' laboratory.}
  \label{fig_plates}
\end{figure}

Borrowing the vocabulary of design by contract~\citep{meyer1992dbc} --- a precondition that must hold before a routine runs, a postcondition it must establish --- a figure type $\tau\in\{\mathrm{diagram},\mathrm{plot},\mathrm{plate}\}$ carries an evidence contract $\kappa(\tau)$:

\begin{itemize}
\item \textbf{Diagram.} Topology, named boxes, and on-figure labels are entailed by $S_{\mathrm{prose}}\cup S_{\mathrm{struct}}$.
A planner proposes labels; we score them against a closed gold inventory. Pixel quality is qualitative.
\item \textbf{Plot.} Labeled data values are entailed by $S_{\mathrm{table}}$, or the figure is marked \emph{illustrative}. Renderer: executable code.
\item \textbf{Plate.} The canvas depicts an in-hand specimen or apparatus under journal multi-panel conventions.
Renderer: observational image generation.
Type and integrity are scored on plate gold ($n=25$); stills appear in Figures~\ref{fig_families} and~\ref{fig_plates}.
\end{itemize}

\begin{figure}[t]
  \centering
  \includegraphics[width=\linewidth]{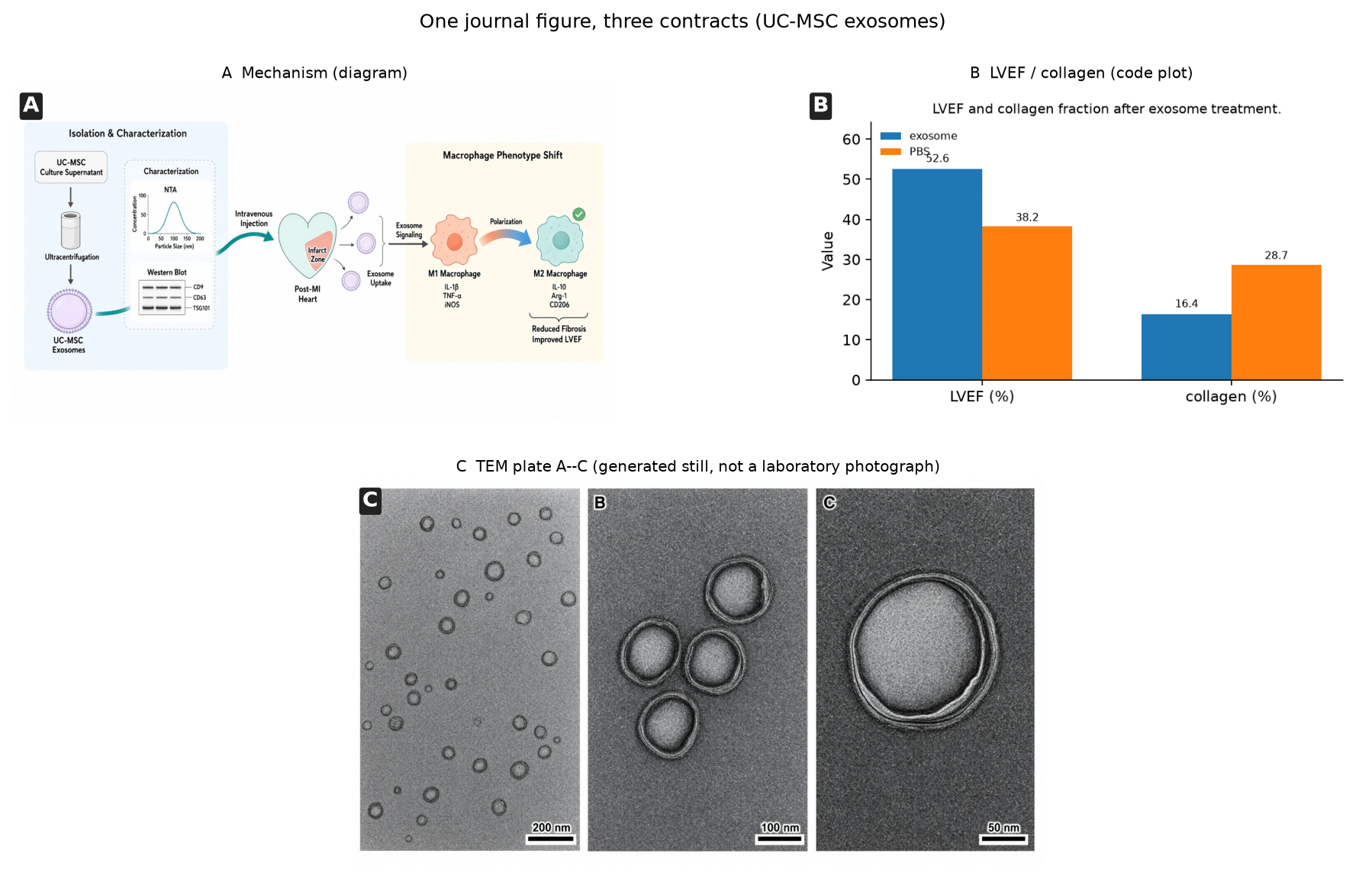}
  \caption{TypedInk composition of one record (UC-MSC exosomes), in the mixed-type layout of Figure~\ref{fig_journal}.
  A~mechanism diagram, B~table-entailed LVEF/collagen (code), C~TEM stills (image template).}
  \label{fig_compose}
\end{figure}

Observational plates dominate empirical science and are absent from NeurIPS-methodology benches~\citep{zhu2026paperbanana,zhu2026autofigure}.
A TEM of cup-shaped vesicles is not a mechanism diagram and is not a bar chart.
Journals print all three contracts under one caption (Figure~\ref{fig_journal}); TypedInk reconstructs that layout from a typed record (Figure~\ref{fig_compose}).

The compiler is
\begin{equation}
\hat\tau = \Pi(S,C), \qquad I = R_{\hat\tau}(S,C).
\label{eq:compiler}
\end{equation}
A \emph{type error} is $\hat\tau\neq\tau^\star$.
An \emph{integrity violation} is a type error that breaks $\kappa(\tau^\star)$: in particular $\tau^\star=\mathrm{plot}$ rendered by an image model, or $\tau^\star=\mathrm{plate}$ forced through a chart grammar.
Tufte's demand that graphics ``show the truth''~\citep{tufte1983vdi,quispel2018aesthetics} becomes, under this formulation, a type-system invariant rather than a style guideline.

Completeness of $S$ is an experimental axis.
We evaluate five views: \texttt{full}, \texttt{struct}, \texttt{prose}, \texttt{table}, and \texttt{intent}.
Gold $\tau^\star$ is the intended family of each constructed illustration, assigned with the target.
It is not defined as $\Pi(S_{\mathrm{full}})$.
\textsc{InferType} is a lexical router that must recover that assignment on the complete record (Section~\ref{sec:experiments}).

\paragraph{Worked example.}
Target \texttt{en\_ml\_retriever\_plot} has gold type \texttt{plot} and table $(0.382, 0.351, 0.291, 0.247)$.
Table~\ref{tab:worked} shows $\Pi$ on each view.
The outline names Recall@10; that is not a table.
The compiler's job on the \texttt{struct} view is to refuse measured ink, not to invent bars.

\begin{table}[t]
\centering
\caption{Worked example: dual-encoder retrieval plot.
Gold $\tau^\star=\mathrm{plot}$.
Integrity fails only when a measured plot would enter an image model.}
\label{tab:worked}
\small
\begin{tabular}{llp{3.4cm}lc}
\toprule
View & Observed $S$ & $\Pi$ & Backend & Integrity \\
\midrule
full   & all fields & plot & code & ok \\
intent & chart intent only & plot & code & ok$^\dagger$ \\
table  & numbers only & plot & code & ok \\
prose  & ``ours versus BM25'' & diagram & image-fast & fail \\
struct & outline of claims & diagram & image-fast & fail \\
\bottomrule
\end{tabular}\\[2pt]
{\raggedright\footnotesize $^\dagger$Typed as plot; with no table the renderer marks the canvas \emph{illustrative} and inks no measured tokens. Backend labels: \texttt{image-fast} is the layout template (Qwen-Image-3.0); \texttt{image-still} is the still template (Qwen-Image-3.0-Pro), used for paper-quality diagram and plate rasters (Section~\ref{sec:method}).\par}
\end{table}

\section{Related Work}
\label{sec:related_work}

\paragraph{Academic diagrams.}
Automated diagram generation is a long-standing challenge~\citep{rodriguez2023figgen}.
Code-based TikZ pipelines produce editable vectors but struggle with the icons and nested layouts of modern AI papers~\citep{belouadi2023automatikz,belouadi2024detikzify}.
Image generators close the visual gap at the cost of editability and of numerical fidelity~\citep{wu2025qwenimage}.
PaperBanana~\citep{zhu2026paperbanana} is the closest agentic baseline: retrieve, plan, style, render, critique, with PaperBananaBench built from $292$ NeurIPS 2025 methodology diagrams (source context avg.\ $3{,}020$ words).
AutoFigure~\citep{zhu2026autofigure} inserts a symbolic blueprint and an OCR-grounded text overlay; FigureBench is $3{,}300$ long-text schematic pairs (research-paper source avg.\ ${>}10$k tokens) and, by construction, drops data-driven charts.
SciFig~\citep{huang2026scifig} maps a method section and caption to editable XML (SciFig-Bench: $435$ author-drawn methodology figures; median method text $1{,}219$ words) and compares raster generators, including Qwen-Image, as single-model baselines.
SridBench~\citep{chang2025sridbench} evaluates illustration drawing across computer science and natural science, still assuming a method-like source.

\paragraph{Statistical plots.}
Plotting from tables is a mature LLM-agent problem~\citep{yang2024matplotagent}.
ChartMimic~\citep{yang2025chartmimic} inverts the map for evaluation.
PaperBanana extends the same agent stack to plots by switching the Visualizer to matplotlib, and separately explores image generation for plots~\citep{zhu2026paperbanana}.
AutoFigure's curation filter is the same distinction stated as a dataset policy rather than as a type system.
SciFig treats programming-based plots as a solved adjacent problem and reports qualitative generalization only.

\paragraph{Protocol comparison.}
Table~\ref{tab:compare} compares published task statements.
PaperBanana, AutoFigure, and SciFig assume a method section or a finished paper whose family is already chosen; we do not rerun their full pipelines end to end (retrieval, planning, styling, critique), since replicating each system's exact orchestration and default backends is out of scope for one preprint.
What all three leave undefined is the case where the source is a fragment rather than a finished section.
If the gold target is a plot and the table is hidden, FigureBench would have excluded the case; PaperBanana's plot Visualizer is an always-draw matplotlib coder, and since PaperBanana's code is public we run its actual Visualizer prompt and execution harness directly as the baseline measured in Section~\ref{sec:experiments} and Table~\ref{tab:compare}'s bottom row (implementation notes in Appendix~\ref{app:naive}).
If the gold target is a TEM plate, none of the three benches treat it as a first-class family.

\begin{table}[t]
\centering
\caption{Protocol comparison with complete-source illustrators.
Published scores stay on their native benches; we do not rerun their full pipelines, except for the bottom row, where we run PaperBanana's actual plot-Visualizer prompt and execution code (Appendix~\ref{app:naive}) on TypedFigBench.
TypedInk's native numbers are type/integrity and refuse-versus-always-ink (Section~\ref{sec:experiments}).}
\label{tab:compare}
\setlength{\tabcolsep}{3.2pt}
\small
\begin{tabularx}{\linewidth}{>{\raggedright\arraybackslash}p{2.05cm}XXXX}
\toprule
 & PaperBanana & AutoFigure & SciFig & TypedInk \\
\midrule
Assumed $S$ & method section $+$ caption & finished document & method section $+$ caption & partial record; fields may be empty \\
Completeness & given & given (long) & given & experimental axis \\
Families & diagram; plot as a Visualizer switch & schematic; charts dropped & method diagram & diagram, plot, plate \\
Infers $\tau$ from $S$? & no (task given) & no & no & $\Pi(S,C)$ \\
Refuses measured ink? & no; image-gen plots explored & N/A (charts excluded) & no (plots are code) & integrity $\neq$ type error \\
Output & raster & layout $+$ overlay & editable XML & code plots; raster diagrams \\
Native bench & PaperBananaBench $292$ & FigureBench $3{,}300$ & SciFig-Bench $435$ & TypedFigBench $16\times 5$ \\
Native metric & VLM 4-axis vs human & $53\%$ paper win vs human & SciFig-Eval, 4 axes & type; integrity; gold recall; refuse \\
\midrule
Hidden-table plot, measured ($n{=}18$, here) & real Visualizer$^\dagger$: $17/18$ inked ($1$ exec.\ fail), gold recall $0.044$ mean when inked, $0/18$ full recovery & N/A (charts excluded) & N/A (plots assumed solved, not evaluated) & $0/18$ inked: $12$ refused, $6$ illustrative \\
\bottomrule
\end{tabularx}\\[2pt]
{\raggedright\footnotesize $^\dagger$PaperBanana's own system prompt, prompt template, and code-execution harness (verbatim from their public repository), run on Qwen3.7-Plus since we lack their default Gemini/OpenAI backends (Appendix~\ref{app:naive}). A separate TypedInk-internal ablation using TypedInk's own prompt (Table~\ref{tab:plots}, naive column) gives $18/18$ inked, gold recall $0.00$.\par}
\end{table}

\section{TypedInk}
\label{sec:method}

\begin{figure}[t]
  \centering
  \begin{tabular}{@{}cc@{}}
    \includegraphics[width=0.48\linewidth]{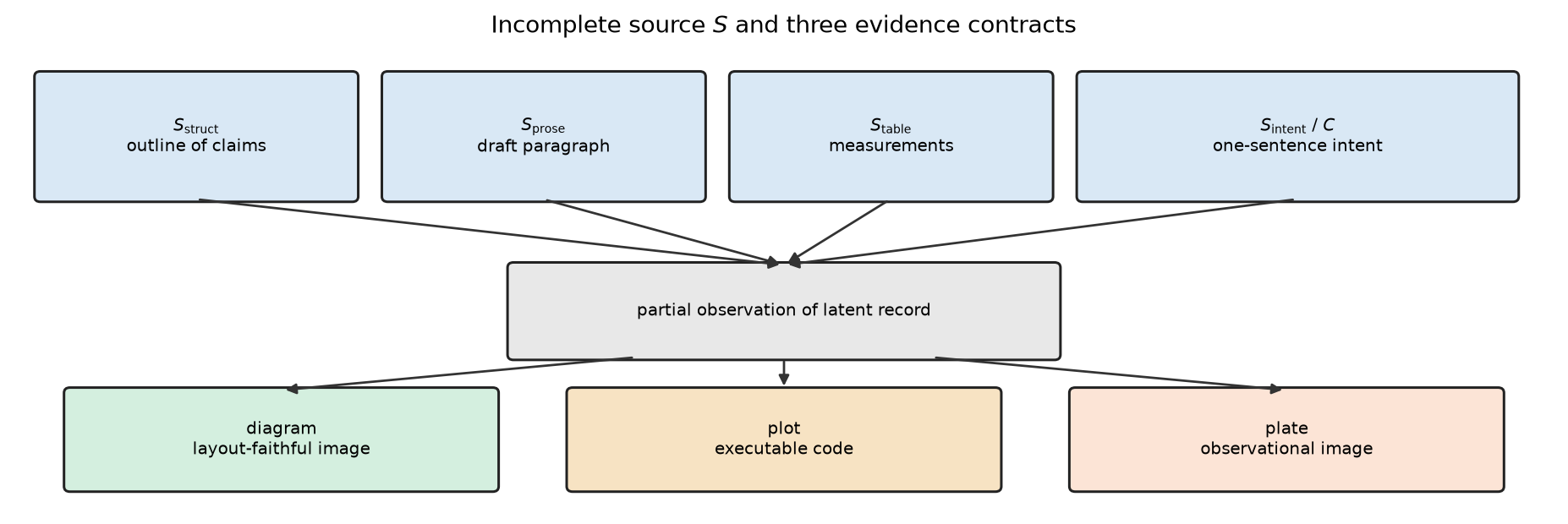} &
    \includegraphics[width=0.48\linewidth]{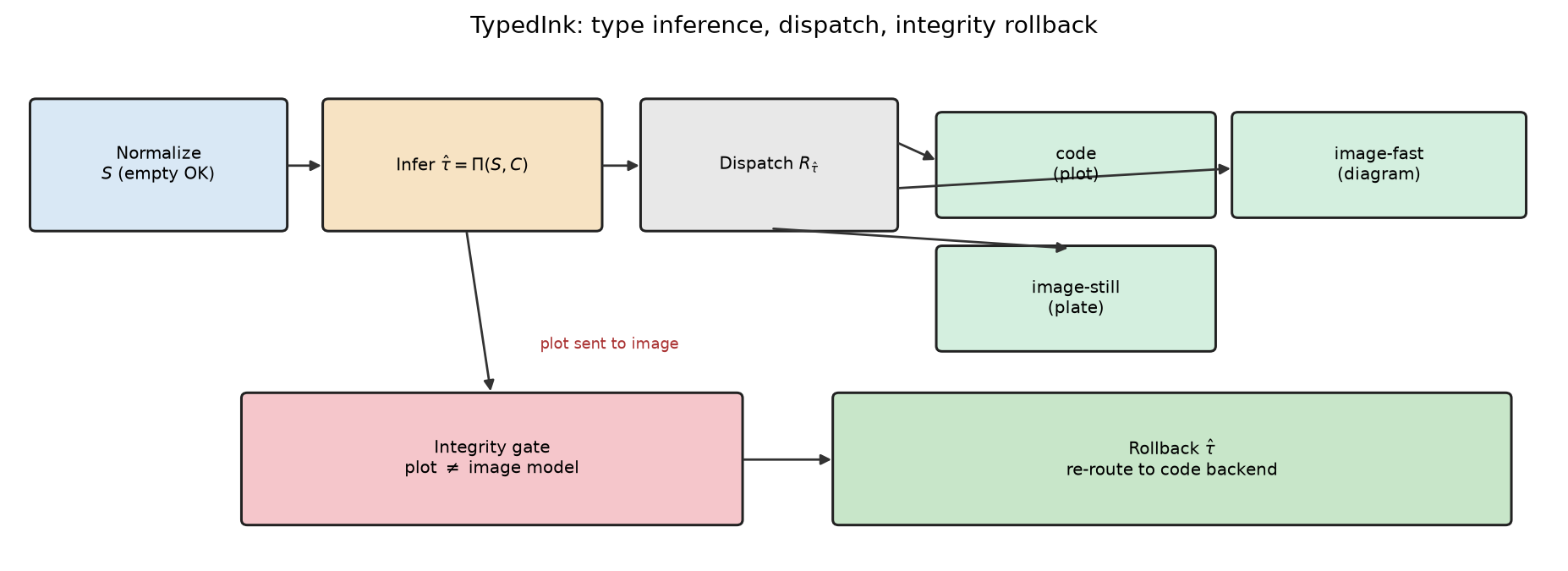} \\
    {\small (a) Incomplete source $S$} & {\small (b) Infer, dispatch, rollback}
  \end{tabular}
  \caption{TypedInk.
  (a)~Any field of $S$ may be empty; diagram, plot, and plate are types, not interchangeable backends.
  (b)~If a gold plot would enter an image model, the integrity gate rolls $\hat\tau$ back to code.}
  \label{fig_method}
\end{figure}

TypedInk is the compiler in Eq.~\eqref{eq:compiler}.
Figure~\ref{fig_method} is the dataflow.
Section~\ref{sec:experiments} evaluates $\Pi$, plot coding against always-draw, diagram labels, plate stills, and a two-rater VL pass on the submit-if-true questions.

\subsection{Normalization}
Whatever arrived is mapped into $S$.
Empty fields stay empty.
The compiler does not impute a methodology section.
If $S_{\mathrm{table}}$ contains fewer than two numeric tokens it is not empirical evidence.

\subsection{Type inference}
Contracts are applied in order by \textsc{InferType}, matching the frozen oracle:

\begin{enumerate}
\item If $C$ names an in-hand specimen or micrograph, and is not simultaneously a chart of tabulated measurements, predict \texttt{plate}.
\item If $S_{\mathrm{table}}$ contains at least two numeric tokens, or $C$ is a chart intent, predict \texttt{plot}.
\item If plate cues remain in the concatenation of nonempty fields (the \emph{blob}) with no table, predict \texttt{plate}.
\item Otherwise predict \texttt{diagram}.
\end{enumerate}

An optional language-model planner may refine captions and layout.
It may not override (1)--(2).
That refusal is the integrity gate acting as a compile gate: illegal dispatch does not become a prettier figure.

\begin{table}[t]
\centering
\small
\begin{tabular}{@{}p{0.94\linewidth}@{}}
\toprule
\textbf{InferType}$(S,C)$ \\
\midrule
\textbf{Input:} source $S=(S_{\mathrm{struct}},S_{\mathrm{prose}},S_{\mathrm{table}},S_{\mathrm{intent}})$, intent $C$ (default $S_{\mathrm{intent}}$).\\
\textbf{Output:} $\hat\tau\in\{\mathrm{diagram},\mathrm{plot},\mathrm{plate}\}$.\\[4pt]
1.\ If $C$ has a plate cue (TEM/SEM/stain/in-hand/\ldots) and is not a measured chart, \textbf{return} \texttt{plate}.\\
2.\ If $S_{\mathrm{table}}$ has $\ge 2$ numeric tokens \textbf{or} $C$ has a chart cue, \textbf{return} \texttt{plot}.\\
3.\ If the concatenation of nonempty fields has a plate cue, \textbf{return} \texttt{plate}.\\
4.\ \textbf{return} \texttt{diagram}.\\
\bottomrule
\end{tabular}
\caption{Deterministic oracle $\Pi$ (\textsc{InferType}).
Chart cues are matched on $C$, not on an outline that merely names a metric.
That is why \texttt{struct} views of gold plots fall through to \texttt{diagram}.}
\label{alg:pi}
\end{table}

\subsection{Typed rendering}
\textbf{Plot.}
A language model writes matplotlib from $S_{\mathrm{table}}$ only after $\Pi$ returns \texttt{plot}; execute; repair on exception (at most three attempts); restyle glyphs by coverage.
No image model.
If the type is plot but the table is empty, the canvas is marked \emph{illustrative} and no measured tokens are inked.
If $\Pi$ does not return plot, the code path is not entered.

\textbf{Diagram.}
A planner proposes a JSON list of on-figure labels and a brief; a layout-oriented text-to-image template paints the page.
Proposed labels are scored against a closed gold inventory $V_\tau$ (precision, recall, extras).
The proposed set comes from the model.

\textbf{Plate.}
A still template paints a white multi-panel page (A--C, scale bar), the layout in Figure~\ref{fig_compose}.
Layout composition and on-figure type are different axes, so $\tau$ selects the template, not only the prompt.
An unpainted lower canvas triggers a square retry.
At most two plates are admitted per paper-length request, and only if the object is in hand.
Type and integrity: Table~\ref{tab:bench}. Stills: Figures~\ref{fig_families} and~\ref{fig_plates}.

Reliability operators---glyph restyle, execution repair, unpainted-canvas retry, prompt-extend gating---are listed in Appendix~\ref{app:protocol}.
Backends in Section~\ref{sec:experiments} are Qwen3.7-Plus~\citep{qwen2025qwen3} (writer), Qwen-Image-3.0~\citep{wu2025qwenimage} (layout), and Qwen-Image-3.0-Pro~\citep{wu2025qwenimage} (stills).

\section{Experiments}
\label{sec:experiments}

TypedFigBench has two layers, and they answer different questions.
Layer A (Sections~\ref{sec:layer_a_start}--\ref{sec:layer_a_end}) is a field ablation of the type router \textsc{InferType} against constructed, author-assigned gold; it diagnoses \emph{why} type inference fails on an incomplete source, and is not by itself a hallucination benchmark, since the author who wrote the router's rules also assigned the gold labels ($\Pi(S_{\mathrm{full}})=\tau^\star$ is a construction check, not a classification result).
Layer B (Section~\ref{sec:layer_b_start} onward) measures hallucinated visual evidence directly, on live generation: gold-table recall and refuse-versus-always-draw on executed matplotlib, plus two-rater VL integrity judgments.
None of Layer B's numbers depend on \textsc{InferType}'s own accuracy.

\subsection{Layer A: a diagnostic field ablation}
\label{sec:layer_a_start}
TypedFigBench is 16 illustrations $\times$ 5 source views ($n=80$): dual-encoder retrieval, tiled attention, exosome isolation, GelMA chondrogenesis, sputter--anneal--section, and a CJK transfer split with the same contracts (Table~\ref{tab:bench}; catalog in Appendix~\ref{app:catalog}).
Gold $\tau^\star$ is the intended family of each target, assigned by the same author who wrote \textsc{InferType}.
The full-view $1.00$ is therefore a construction check that $\Pi(S_{\mathrm{full}})=\tau^\star$, not an independent classification result; the ablation's value is in the other four views, where the router's rules and the gold labels can disagree.
Metrics are type accuracy and integrity-fail rate.

\begin{table}[t]
\centering
\caption{TypedFigBench inventory (top), Layer~A by source view (middle), and by gold type (bottom).
Full-view $1.00$ is $\Pi(S_{\mathrm{full}})=\tau^\star$.
Plate gold ($n=25$): type and integrity.}
\label{tab:bench}
\small
\begin{tabular}{lcccc}
\toprule
Domain & $n$ & diagram & plot & plate \\
\midrule
ML            & 4 & 2 & 2 & 0 \\
biomedicine   & 8 & 2 & 3 & 3 \\
materials     & 4 & 1 & 1 & 2 \\
\midrule
total         & 16 & 5 & 6 & 5 \\
\bottomrule
\end{tabular}
\vspace{0.9em}

\begin{tabular}{lccc}
\toprule
View & $n$ & Type acc. & Integrity fail \\
\midrule
full   & 16 & 1.00 & 0.00 \\
intent & 16 & 1.00 & 0.00 \\
table  & 16 & 0.69 & 0.00 \\
prose  & 16 & 0.56 & 0.38 \\
struct & 16 & 0.31 & 0.38 \\
\midrule
overall & 80 & 0.71 & 0.15 \\
\bottomrule
\end{tabular}
\vspace{0.9em}

\begin{tabular}{lccc}
\toprule
Gold $\tau^\star$ & $n$ & Type acc. & Integrity fail \\
\midrule
diagram & 25 & 1.00 & 0.00 \\
plot    & 30 & 0.60 & 0.40 \\
plate   & 25 & 0.56 & 0.00 \\
\bottomrule
\end{tabular}
\end{table}

\subsection{Which fields the specified router reads}
Layer~A is a field ablation of \textsc{InferType} (Table~\ref{tab:bench}, Figure~\ref{fig_layer_a}).
The complete record matches gold; a one-sentence intent also matches, because $C$ names the family.
A table never sends a measured plot to an image model (integrity-fail $0.00$), but type accuracy is only $0.69$: empty tables of diagrams and plates collapse to \texttt{diagram}.
Prose and structure are the dangerous views: type accuracy $0.56$ and $0.31$, integrity-fail $0.375$.
An outline is not a scientific record under these rules---which is why the compiler has a gate, and why always-draw still matters.

\subsection{Plots, not plates, break integrity}
\begin{figure}[t]
  \centering
  \includegraphics[width=\linewidth]{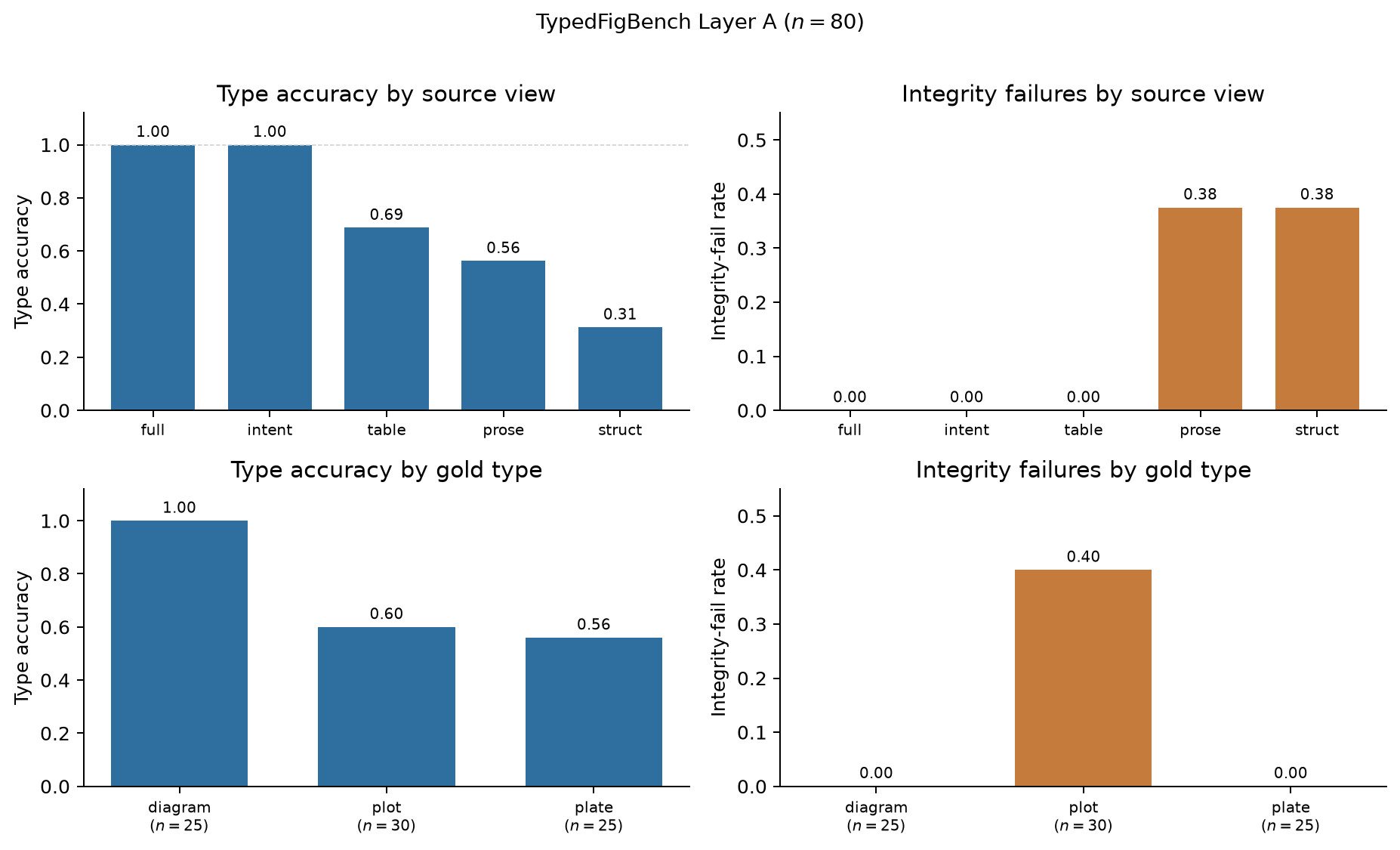}
  \caption{Layer~A ($n=80$).
  Integrity failures concentrate on plots whose table is hidden ($0.40$).
  Gold plates: type accuracy $0.56$, integrity-fail $0.00$.}
  \label{fig_layer_a}
\end{figure}

Figure~\ref{fig_layer_a} (bottom) and Table~\ref{tab:bench} (bottom) split the same 80 cases by gold type.
Diagrams are recovered under every view that still contains mechanism language (type accuracy $1.00$, integrity-fail $0.00$).
Plates are the quantitative third family at this layer: $n=25$, type accuracy $0.56$, integrity-fail $0.00$.
They are missed when the micrograph cue is stripped, but they are never forced through a chart grammar.
Gold plots have type accuracy $0.60$ and integrity-fail $0.40$: when $S_{\mathrm{table}}$ is absent, $\Pi$ falls through to \texttt{diagram}.
On-figure names in $S_{\mathrm{prose}}\cup S_{\mathrm{struct}}$ cover $1.00$ of the gold inventory on the full record, $0.32$ on the outline, and $0$ on intent or table.

Table~\ref{tab:errors} lists every integrity failure.
They are exactly the six gold plots under \texttt{struct} and \texttt{prose} ($12$ of $12$ such views).
Two biomedicine plot-\texttt{prose} views are predicted \texttt{plate} rather than \texttt{diagram}, because the paragraph contains Masson/TEM language.
These still count as integrity violations, since a still-image backend would end up painting LVEF.
By contrast, plate type errors on \texttt{struct} and empty \texttt{table} are misses rather than integrity failures: the compiler draws a diagram instead of a micrograph, but it never invents a chart of the specimen.

Domain splits are secondary (ML $0.80$ type acc., biomedicine $0.68$, materials $0.70$).
The CJK split uses the same contracts; plate cues remain lexical.

\begin{table}[t]
\centering
\caption{Every Layer~A integrity failure ($n=12$).
All are gold plots under \texttt{struct} or \texttt{prose}.
No diagram or plate row appears.}
\label{tab:errors}
\small
\begin{tabular}{lllc}
\toprule
Target & View & $\Pi$ & Would dispatch \\
\midrule
\texttt{en\_ml\_retriever\_plot} & struct, prose & diagram & image-fast \\
\texttt{en\_ml\_attn\_plot} & struct, prose & diagram & image-fast \\
\texttt{en\_bio\_exosome\_plot} & struct & diagram & image-fast \\
\texttt{en\_bio\_exosome\_plot} & prose & plate & image-still \\
\texttt{en\_bio\_gelma\_plot} & struct, prose & diagram & image-fast \\
\texttt{en\_mat\_modulus\_plot} & struct, prose & diagram & image-fast \\
\texttt{zh\_bio\_exosome\_plot} & struct & diagram & image-fast \\
\texttt{zh\_bio\_exosome\_plot} & prose & plate & image-still \\
\bottomrule
\end{tabular}
\end{table}

\subsection{Lexical oracle versus a chat-model planner}
\label{sec:layer_a_end}
\textsc{InferType} is keyword cues on $C$ plus a numeric-token test on $S_{\mathrm{table}}$.
A chat-model planner (Qwen3.7-Plus, parse-fail $0$) sees the same ablated $S$ and is scored against the same gold (Table~\ref{tab:llm_type}).
Type accuracy rises from $0.71$ to $0.85$ and integrity-fail falls from $0.15$ to $0.05$, mostly on prose ($0.56\rightarrow 0.94$).
Outlines remain dangerous: $4/16$ struct cases still integrity-fail, including two gold plates predicted \texttt{plot}.
The planner uses more than lexical cues; the gate still has work to do.

\begin{table}[t]
\centering
\caption{Type recovery vs.\ author-assigned gold.
Top: by source view ($n=16$ per view).
Bottom: by gold type (plates $n=25$).}
\label{tab:llm_type}
\small
\begin{tabular}{lcccc}
\toprule
View & Oracle acc. & LLM acc. & Oracle fail & LLM fail \\
\midrule
full   & 1.00 & 1.00 & 0.00 & 0.00 \\
intent & 1.00 & 1.00 & 0.00 & 0.00 \\
table  & 0.69 & 0.69 & 0.00 & 0.00 \\
prose  & 0.56 & 0.94 & 0.38 & 0.00 \\
struct & 0.31 & 0.63 & 0.38 & 0.25 \\
\midrule
overall & 0.71 & 0.85 & 0.15 & 0.05 \\
\bottomrule
\end{tabular}
\vspace{0.9em}

\begin{tabular}{lccccc}
\toprule
Gold $\tau^\star$ & $n$ & Oracle acc. & LLM acc. & Oracle fail & LLM fail \\
\midrule
diagram & 25 & 1.00 & 0.88 & 0.00 & 0.00 \\
plot    & 30 & 0.60 & 0.93 & 0.40 & 0.07 \\
plate   & 25 & 0.56 & 0.72 & 0.00 & 0.08 \\
\bottomrule
\end{tabular}
\end{table}

\subsection{Layer B: does TypedInk hallucinate on live generation?}
\label{sec:layer_b_start}
The six gold plots $\times$ five views ($n=30$) are the integrity experiment that Layer~A only diagnoses.
Unlike Layer~A, every number in this subsection and the two that follow comes from executed code or a live VL rating, not from \textsc{InferType}'s own type-accuracy table.
TypedInk runs $\Pi$, then asks the chat model to write matplotlib from $S_{\mathrm{table}}$ only.
An untyped baseline skips $\Pi$ and writes matplotlib from the entire observed blob, holding TypedInk's own prompt and model fixed so the type gate is the only manipulated variable (Table~\ref{tab:plots}); a second, independent baseline that runs PaperBanana's own public Visualizer code directly gives a consistent result (Table~\ref{tab:compare}, Appendix~\ref{app:naive}).

When a table is present (\texttt{full} and \texttt{table}, $n=12$), both paths execute and both recover every gold number string (gold recall $1.00$, execution failures $0$).
The paths diverge when the table is hidden.
TypedInk refuses \texttt{struct} and \texttt{prose} ($12/12$) and marks \texttt{intent} illustrative ($6/6$, no measured ink).
The untyped path inks $30/30$ charts, including all $18$ hidden-table views.
Gold recall on those $18$ is $0.00$: the model does not copy the gold table that was never shown.
It still draws a chart.
Figure~\ref{fig_live}(b) is one such chart: an outline that names Recall@10 becomes a three-method bar graph with invented heights.
Token-level Jaccard against the whole matplotlib script is not a finding (figsize and dpi dominate); gold recall and refuse-versus-always-ink are.

\begin{figure}[t]
  \centering
  \begin{tabular}{@{}cc@{}}
    \includegraphics[width=0.48\linewidth]{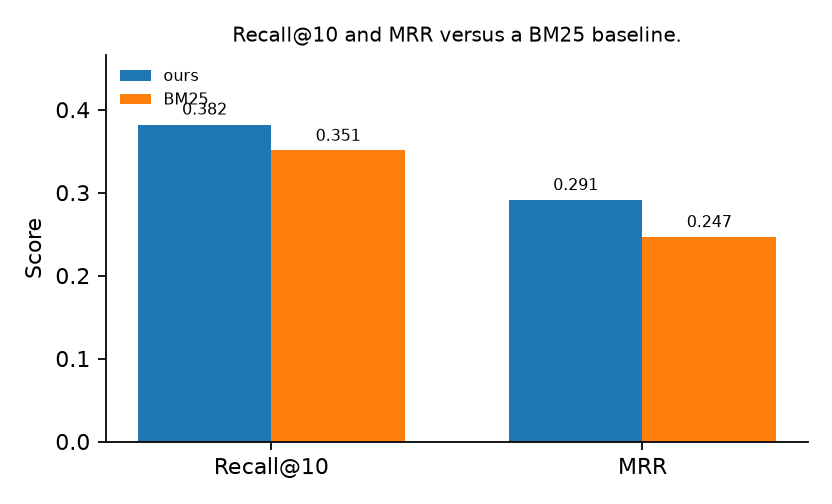} &
    \includegraphics[width=0.48\linewidth]{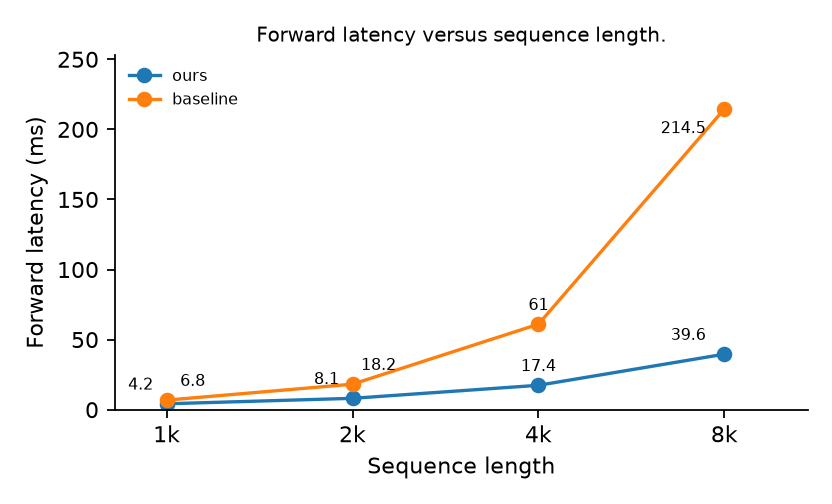} \\
    {\small (a) Retrieval vs BM25} & {\small (b) Tiled-attention latency} \\[4pt]
    \includegraphics[width=0.48\linewidth]{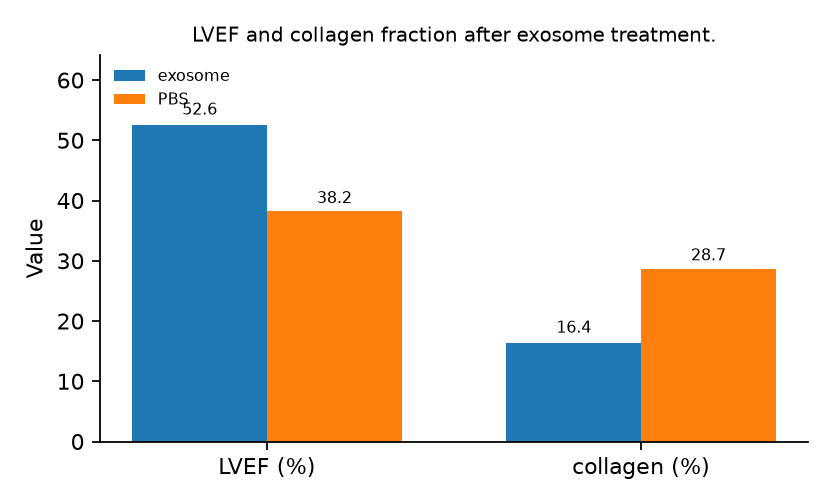} &
    \includegraphics[width=0.48\linewidth]{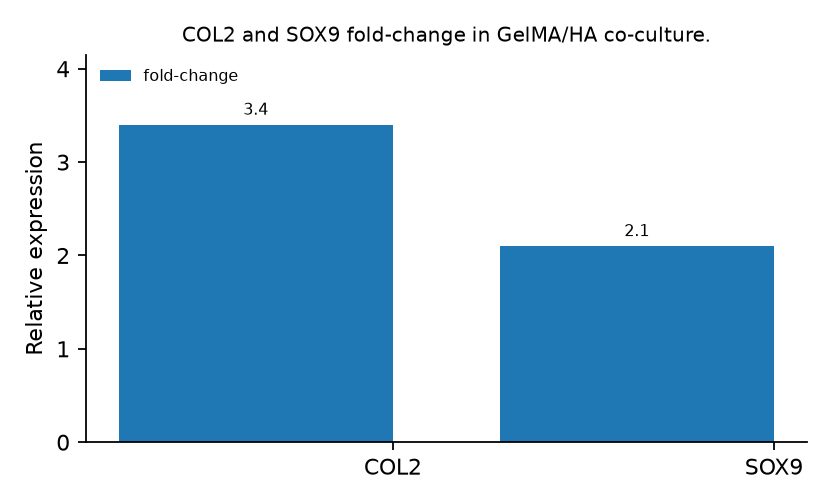} \\
    {\small (c) LVEF and collagen} & {\small (d) COL2 / SOX9 fold-change} \\[4pt]
    \includegraphics[width=0.48\linewidth]{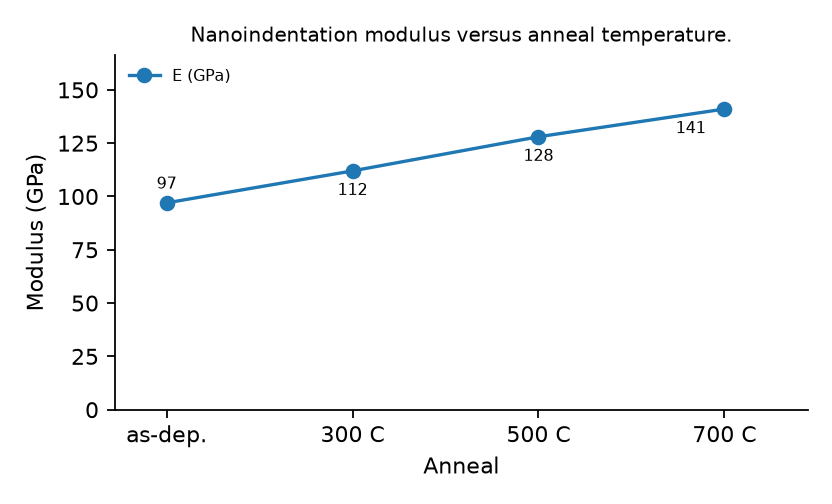} &
    \includegraphics[width=0.48\linewidth]{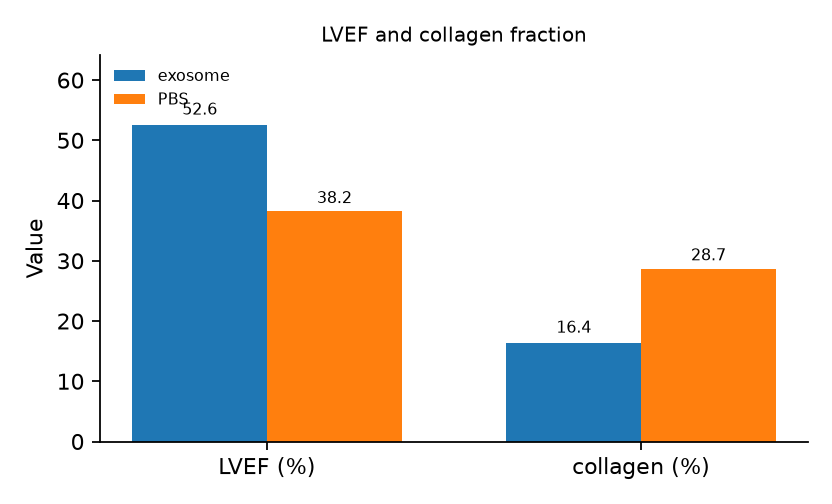} \\
    {\small (e) Nanoindentation modulus} & {\small (f) CJK transfer, same table}
  \end{tabular}
  \caption{Contract illustration: gold-spec code-backend plots from $S_{\mathrm{table}}$ (full view).
  Every labeled number appears in the corresponding gold table.
  These are not image-model outputs.}
  \label{fig_code}
\end{figure}

Gold-spec plots from the table (Figure~\ref{fig_code}) remain the contract illustration: every labeled bar height is in $S_{\mathrm{table}}$ by construction.
Figure~\ref{fig_live}(a) is the live coder on a measured table.

\begin{table}[t]
\centering
\caption{Six gold plots $\times$ five views ($n=30$).
TypedInk: $\Pi$ then LLM matplotlib from $S_{\mathrm{table}}$.
Naive: LLM matplotlib from the whole observed blob, no type gate.
Gold recall is the fraction of gold table number-strings that appear in executed code, averaged over inked rows.}
\label{tab:plots}
\small
\begin{tabular}{lcccc}
\toprule
View & TypedInk & Naive & Gold rec.\ (TI) & Gold rec.\ (naive) \\
\midrule
full   & 6/6 inked & 6/6 inked & 1.00 & 1.00 \\
table  & 6/6 inked & 6/6 inked & 1.00 & 1.00 \\
intent & 6/6 illustrative & 6/6 inked & --- & 0.00 \\
prose  & 6/6 refused & 6/6 inked & --- & 0.00 \\
struct & 6/6 refused & 6/6 inked & --- & 0.00 \\
\bottomrule
\end{tabular}
\end{table}

\subsection{Diagram labels}
\begin{figure}[t]
  \centering
  \begin{tabular}{@{}cc@{}}
    \includegraphics[width=0.48\linewidth]{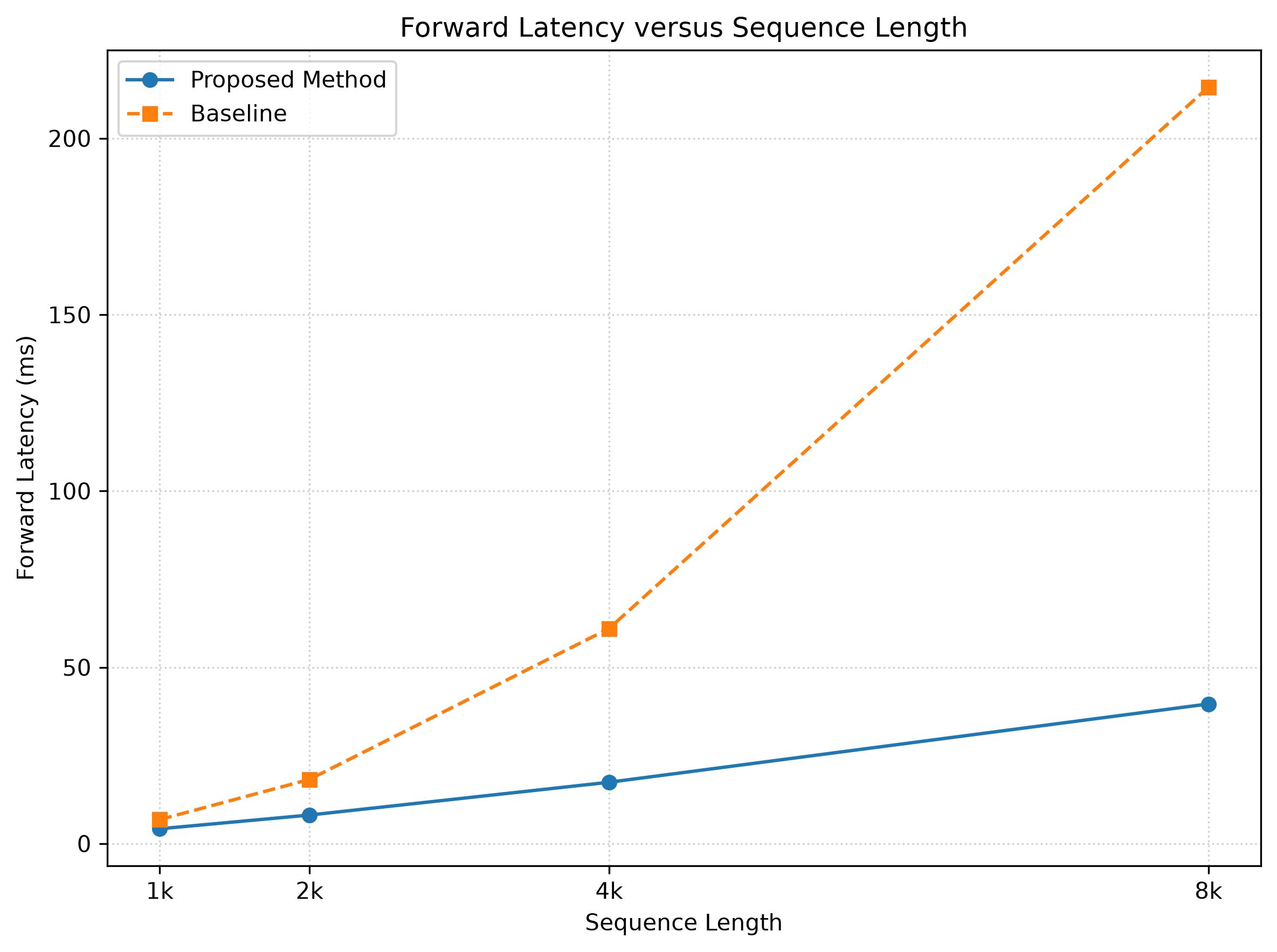} &
    \includegraphics[width=0.48\linewidth]{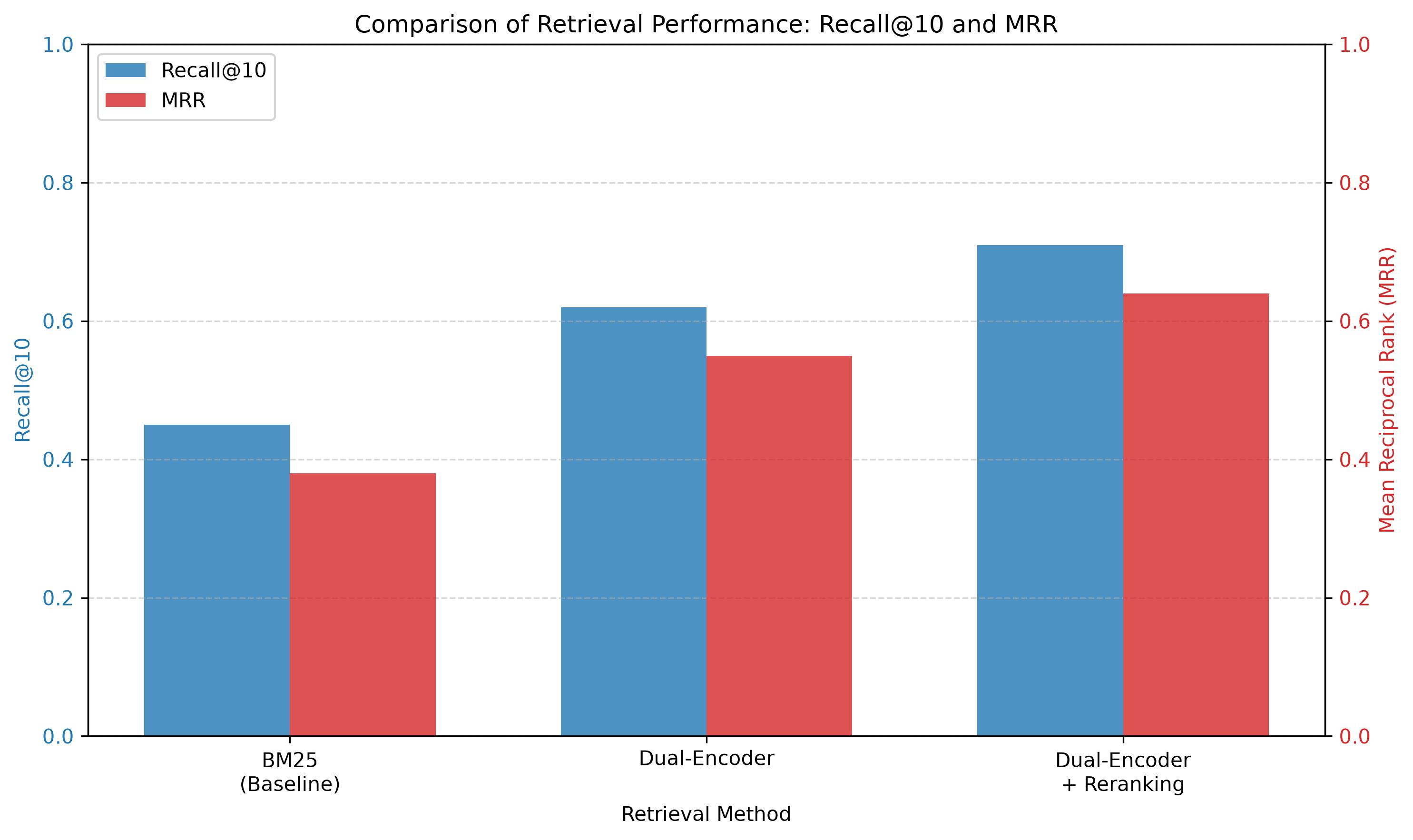} \\
    {\small (a) TypedInk, full (code)} & {\small (b) Untyped, struct (code)} \\[4pt]
    \includegraphics[width=0.48\linewidth]{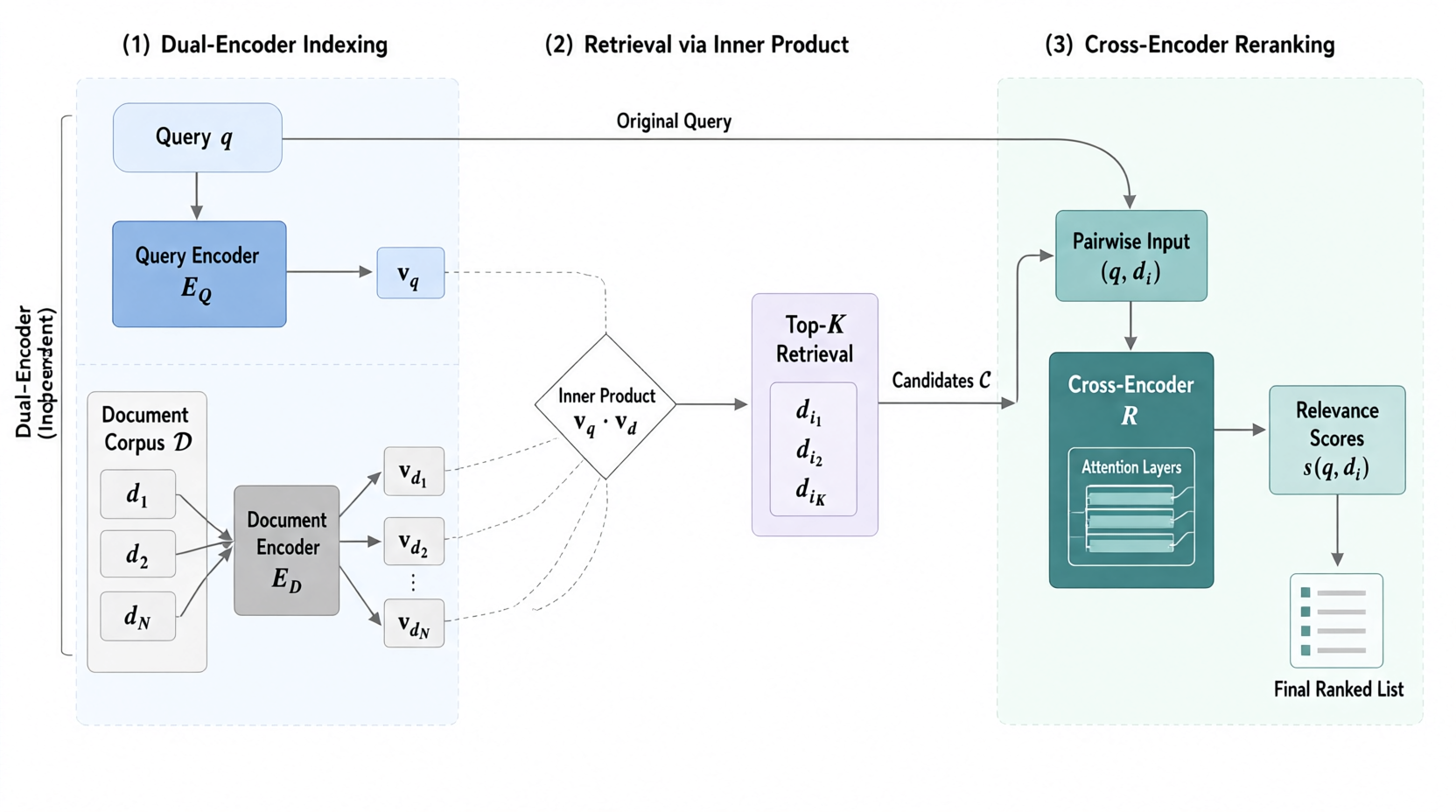} &
    \includegraphics[width=0.48\linewidth]{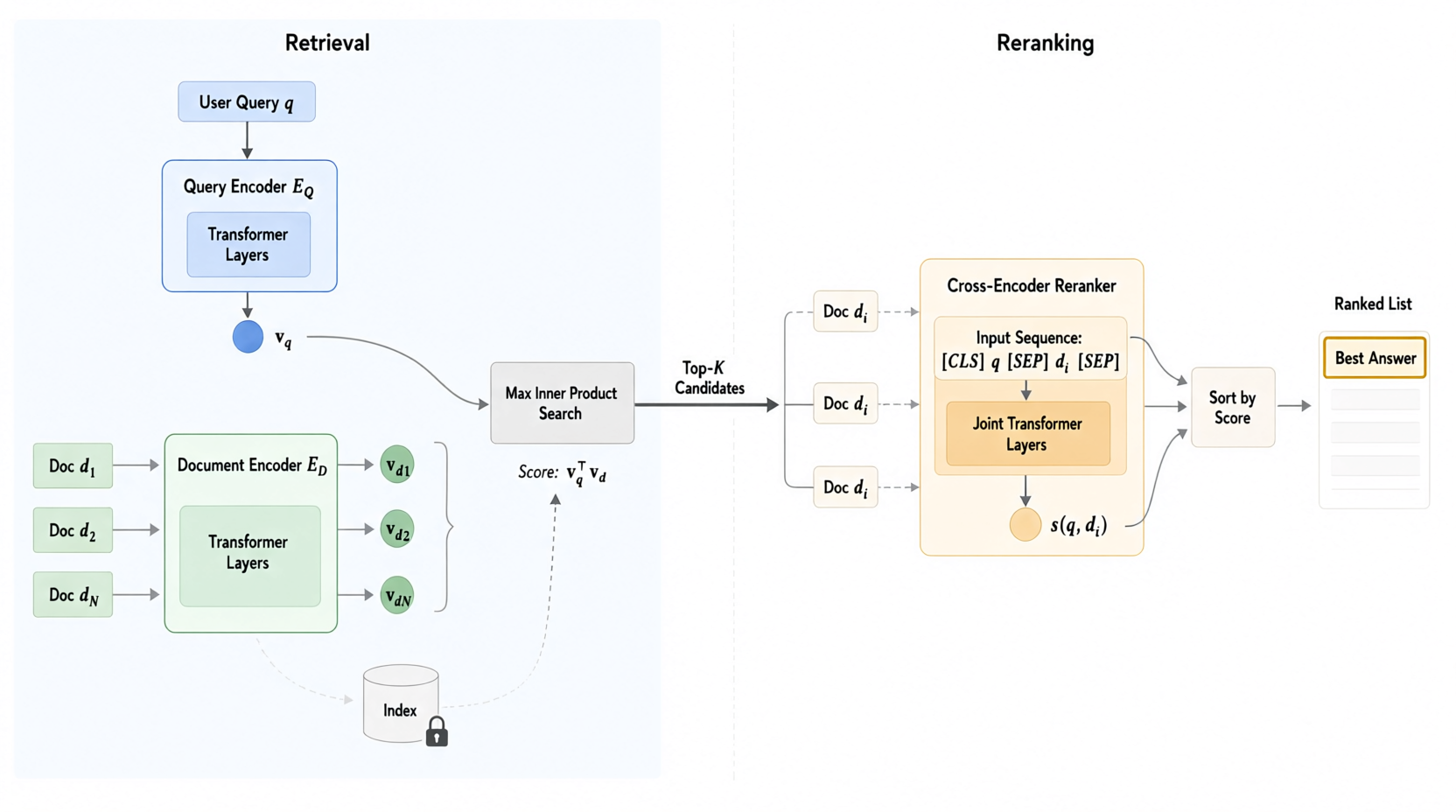} \\
    {\small (c) Diagram, full (still template)} & {\small (d) Diagram, struct (still template)}
  \end{tabular}
  \caption{Live generation.
  (a)~TypedInk writes matplotlib from a measured table.
  (b)~The same coder, given only a retrieval outline, still draws a chart.
  (c--d)~Still-template diagrams: dual-encoder then reranker, full record vs.\ outline.}
  \label{fig_live}
\end{figure}

Five gold diagrams $\times$ five views ($n=25$): the planner proposes JSON labels against a closed inventory $V_\tau$ (Table~\ref{tab:diagram}).
Mean precision $0.54$, mean recall $0.27$.
Full-record recall $0.47$, outline $0.19$, table $0.00$.
Figure~\ref{fig_live}(c--d) renders the still template: (c) keeps dual-encoder indexing, inner-product retrieval, and cross-encoder reranking; (d) is the same pipeline from an outline, with fewer named boxes.

\begin{table}[t]
\centering
\caption{Diagram label proposal vs.\ closed gold inventory $V_\tau$ ($n=5$ per view).
Proposed names come from the planner.}
\label{tab:diagram}
\small
\begin{tabular}{lccc}
\toprule
View & Precision & Recall & Extras / fig. \\
\midrule
full   & 0.53 & 0.47 & 2.6 \\
prose  & 0.56 & 0.44 & 2.2 \\
intent & 0.50 & 0.24 & 1.6 \\
struct & 0.56 & 0.19 & 0.8 \\
table  & --- & 0.00 & 0.0 \\
\midrule
overall & 0.54 & 0.27 & --- \\
\bottomrule
\end{tabular}
\end{table}

\subsection{Why the gate is not a pixel judge}
\label{sec:judge}
Two vision-language raters (Qwen-VL-Plus, Qwen-VL-Max~\citep{qwen2025qwen3vl}) scored TypedInk against always-draw on the same 30 plot pairs, and scored the six plate stills on the photograph question (Table~\ref{tab:judge}).
A/B order was hashed per item.
Cohen's $\kappa$~\citep{cohen1960kappa} on plot submit-if-true is $0.49$.

On hidden-table views ($n=36$ ratings) they never picked the refusal card ($0/36$).
Always-draw was chosen $17/36$ times; neither $19/36$.
On measured full/table views ($n=24$), TypedInk $10$, always-draw $12$, both $2$.
All twelve plate ratings mark the generated stills as laboratory photographs.
Pixel judges still like a chart, and they do not police the specimen contract.
TypedInk therefore refuses before paint.

\begin{table}[t]
\centering
\caption{Submit-if-true and the photograph question (two VL raters).
Plot: TypedInk vs.\ always-draw.
Plate: stills in Figure~\ref{fig_plates}.}
\label{tab:judge}
\small
\begin{tabular}{llp{5.2cm}}
\toprule
Question & $n$ & Outcome \\
\midrule
Plot submit, hidden table & 36 & TypedInk 0, always-draw 17, neither 19 \\
Plot submit, full/table & 24 & TypedInk 10, always-draw 12, both 2 \\
Plate as lab photograph & 12 & 12 yes \\
Plate as generated still & 12 & 12 no \\
\bottomrule
\end{tabular}
\end{table}

\section{Discussion}
\label{sec:discussion}

\paragraph{Outlines carry no type information.}
The struct view of a dual-encoder paper mentions Recall@10 in an experiments heading, but that string is not $S_{\mathrm{table}}$, so \textsc{InferType} does not return \texttt{plot} and the integrity predicate fires.
A complete-document illustrator never encounters this case, since AutoFigure's input is already a finished paper with a chosen conceptual figure~\citep{zhu2026autofigure}.
Skipping the gate produces the untyped coder's behavior in Figure~\ref{fig_live}(b): a plausible results figure with no measured evidence.

\paragraph{Intent is not evidence.}
Communicative intent matches gold under \textsc{InferType} because the constructed $C$ names the family, but naming the family does not license measured ink.
The \texttt{intent} view of every gold plot is dispatched to code and then marked illustrative, since its table is empty, whereas the untyped path inks that view anyway (gold recall $0.00$).
That gap is the difference between a caption and a result.

\paragraph{Plates fail differently.}
Hiding the micrograph cue yields a type error (diagram instead of plate) rather than an integrity violation.
The dangerous confusion for plates --- forcing them through a chart grammar --- never occurs under the lexical oracle, though the chat-model planner does make that mistake on two struct plates (predicted \texttt{plot}), so the gate still has work to do there.
The remaining plate errors are recall failures, not hallucinated numbers.

\paragraph{The gate is earlier than the judge.}
A pretty bar chart of invented LVEF can still score well on aesthetics~\citep{zhu2026autofigure}.
Table~\ref{tab:plots} shows the code-level version of this problem: $18/18$ hidden-table views still leave a chart on the page.
Table~\ref{tab:judge} shows the pixel-level version: VL raters still pick always-draw $17/36$ times and accept generated plates as photographs.
TypedInk's response is to measure refuse-versus-always-ink rather than a four-axis win-rate on method-section benches (Table~\ref{tab:compare}).

\section{Limitations, Risks, and Next Experiments}
\label{sec:limitations}

TypedFigBench is 16 author-built targets; gold type is assigned with each illustration, and $\Pi(S_{\mathrm{full}})=\tau^\star$ is a construction check.
The always-draw result does not rest on that taxonomy: a hidden table either is or is not in $S$.
Output of $R_{\mathrm{diagram}}$ and $R_{\mathrm{plate}}$ is raster, not SciFig XML.
Appendix~\ref{app:protocol} specifies a PaperBananaBench visualizer swap, a backend Pareto, and four-axis aesthetics.

Observational stills in Figures~\ref{fig_families},~\ref{fig_compose}, and~\ref{fig_plates} are generated, not laboratory photographs; Figure~\ref{fig_journal} alone is a CC-BY reprint, and Figure~\ref{fig_compose} recomposes TypedInk's layout on a different record.
Two VL raters mistake those stills for laboratory photographs in every trial (Table~\ref{tab:judge}).
Plate type and integrity remain the primary plate metric ($n=25$, oracle integrity-fail $0.00$); the dual-human-annotation protocol in Appendix~\ref{app:protocol} is unrun.

\paragraph{Dual-use risk.}
The plate result above is not only evidence that pixel judges are unreliable: it is evidence that the still-image backend TypedInk calls can produce fabricated experimental documentation convincing enough that two independent VL raters accepted it as a genuine laboratory photograph on all twelve trials.
TypedInk's integrity gate constrains \emph{when} that backend is invoked inside this pipeline --- only for an in-hand specimen, capped at two plates per request --- but it does not, and cannot, stop the same backend from being called directly, outside TypedInk, to generate a plate and present it as real data.
Caption-level generated/real disclosure, which we enforce on every still in this paper (Figures~\ref{fig_families},~\ref{fig_compose}, and~\ref{fig_plates}), is a norm TypedInk's own output follows, not a technical barrier against someone else misusing the same underlying model.
We surface this explicitly rather than let the judge-unreliability framing above stand alone.

The planner may not override the plot and in-hand contracts; it still integrity-fails on a quarter of outlines.

\section{Conclusion}
\label{sec:conclusion}

TypedInk is a compiler for scientific illustration from incomplete evidence.
It infers diagram, plot, or plate, keeps the code path closed until a table is observed, inks gold numbers when it is, and treats plates as a family at dispatch.
An outline is not enough to recover type; a hidden table is what turns a measured plot into an always-draw hallucination.
That hallucination is measured directly on live generation --- gold-table recall, refuse-versus-always-draw, and PaperBanana's own Visualizer code and VL-judged integrity all agree --- not inferred from the 16-illustration field ablation, which only diagnoses where the type router itself goes wrong.

\appendix
\section{TypedFigBench catalog}
\label{app:catalog}

Table~\ref{tab:catalog} lists the 16 constructed illustrations.
Captions are the communicative targets; gold $\tau^\star$ is the intended family of each target, assigned with the example.
CJK rows use the same contracts as the English biomedicine and materials plates.

\begin{table}[h]
\centering
\caption{TypedFigBench constructed targets. $\tau^\star$ is author-assigned intended family.}
\label{tab:catalog}
\small
\begin{tabular}{lllp{5.6cm}}
\toprule
ID & Domain & $\tau^\star$ & Caption \\
\midrule
\texttt{en\_ml\_retriever\_diagram} & ML & diagram & Dual-encoder retrieval with reranker \\
\texttt{en\_ml\_retriever\_plot} & ML & plot & Recall@10 and MRR vs BM25 \\
\texttt{en\_ml\_attn\_diagram} & ML & diagram & Tiled attention dataflow \\
\texttt{en\_ml\_attn\_plot} & ML & plot & Forward latency vs sequence length \\
\texttt{en\_bio\_exosome\_diagram} & biomed & diagram & Isolation and M1/M2 polarization \\
\texttt{en\_bio\_exosome\_plot} & biomed & plot & LVEF and collagen vs PBS \\
\texttt{en\_bio\_exosome\_plate} & biomed & plate & TEM of cup-shaped vesicles \\
\texttt{en\_bio\_gelma\_plot} & biomed & plot & COL2 and SOX9 fold-change \\
\texttt{en\_bio\_gelma\_plate} & biomed & plate & Safranin O histology, 12 weeks \\
\texttt{en\_mat\_interface\_diagram} & materials & diagram & Sputter, anneal, section \\
\texttt{en\_mat\_modulus\_plot} & materials & plot & Modulus vs anneal temperature \\
\texttt{en\_mat\_sem\_plate} & materials & plate & Cross-sectional SEM, panels A--C \\
\texttt{zh\_bio\_exosome\_diagram} & biomed & diagram & Isolation and polarization (CJK) \\
\texttt{zh\_bio\_exosome\_plot} & biomed & plot & LVEF and collagen (CJK) \\
\texttt{zh\_bio\_exosome\_plate} & biomed & plate & TEM plate (CJK) \\
\texttt{zh\_mat\_sem\_plate} & materials & plate & SEM plate (CJK) \\
\bottomrule
\end{tabular}
\end{table}

\section{Layer B protocol}
\label{app:protocol}

Items marked run are reported in Section~\ref{sec:experiments}.
The rest are specified so a conference revision does not invent a protocol.

\begin{enumerate}
\item \textbf{Source-ablation renders (partially run).}
Same 16 targets, all five views.
Plots: LLM matplotlib with and without $\Pi$ (run; gold recall and refuse-versus-always-ink).
Diagrams: planner-proposed labels vs.\ $V_\tau$ (run); eight English rasters, qualitative (run).
Plates: type and integrity on all 25 gold-plate views (run; Table~\ref{tab:bench}, Table~\ref{tab:llm_type}).
Qualitative still-template rasters (TEM, Safranin O, SEM) in Figures~\ref{fig_families} and~\ref{fig_plates} (run; not laboratory photographs).
VLM-as-judge four-axis aesthetics following~\citet{zhu2026paperbanana} as a bake-off: not run.
Integrity questions (submit-if-true / submit-as-photograph): run as two VL raters (item 5).
\item \textbf{Transfer (unrun).} Keep PaperBananaBench retrieval and planning; swap only the visualizer (vanilla image model vs TypedInk diagram path) on a 50-diagram split.
\item \textbf{Pareto (unrun).} All-fast-image / all-still-image / all-code / TypedInk on wall-clock and API cost.
\item \textbf{Operators (unrun).} $\pm$ prompt-extend gate, $\pm$ unpainted-canvas retry, $\pm$ glyph restyle, $\pm$ few-shot.
\item \textbf{Integrity questions (VL raters).}
Two independent vision-language raters (Qwen-VL-Plus, Qwen-VL-Max).
Thirty plot pairs plus six plate stills (Table~\ref{tab:judge}).
Four PaperBanana axes are collected on plot pairs; the binary questions below are the integrity items.
\begin{itemize}
\item \emph{Plot integrity} (30 pairs, run).
TypedInk vs.\ always-draw on the same gold-plot view; A/B order hashed.
Binary: ``Would you put this in a submission if the numbers must be true?''
\item \emph{Plate specimen} (6 stills, run).
Still-template TEM, Safranin~O, and SEM rasters (two samples each of \texttt{en\_bio\_exosome\_plate}, \texttt{en\_bio\_gelma\_plate}, \texttt{en\_mat\_sem\_plate}).
Two binaries: ``Would you submit this panel as a laboratory photograph of the named specimen?'' and ``Would you submit it as a generated illustration under a caption that says so?''
\end{itemize}
Cohen's $\kappa$ is reported between the two VL raters; dual human annotation of the same sheet is unrun.
\end{enumerate}

\section{Two hidden-table baselines: implementation notes}
\label{app:naive}

We report two different ``no type gate'' baselines on the same 18 hidden-table views, and they answer different questions.

\paragraph{TypedInk-internal ablation (Table~\ref{tab:plots}, naive column).}
This baseline calls the identical code-generation routine as TypedInk's own typed plot path --- same writer model (Qwen3.7-Plus), same system prompt, the same execute-and-repair loop (at most three attempts) --- so it differs from TypedInk in exactly one variable: TypedInk is gated behind $\Pi(S,C)=\mathrm{plot}$ and prompted with $S_{\mathrm{table}}$ only, while this baseline is ungated and prompted with the caption plus the full formatted source for that view. No lint or type check runs on its output before scoring. This isolates the type gate as the only manipulated variable, holding TypedInk's own prompt engineering fixed; it says nothing about PaperBanana specifically.

\paragraph{PaperBanana plot Visualizer, verbatim (Table~\ref{tab:compare}, bottom row).}
PaperBanana's code is public (\url{https://github.com/dwzhu-pku/PaperBanana}, Apache-2.0); we cloned it and confirmed its plot-Visualizer agent (\texttt{agents/visualizer\_agent.py}) uses a fixed system prompt (``You are an expert statistical plot illustrator. Write code to generate high-quality statistical plots based on user requests.''), a fixed prompt template (``Use python matplotlib to generate a statistical plot based on the following detailed description: \{desc\}\textbackslash n Only provide the code without any explanations. Code:''), and a standalone code-execution helper (\texttt{utils/plot\_execution.py}, \texttt{execute\_plot\_code\_worker}) with no import restrictions and no type gate.
We copied these three artifacts verbatim and ran them on the same 18 hidden-table views, substituting Qwen3.7-Plus for PaperBanana's default Gemini/OpenAI/Anthropic backends, since we do not have API access to those; \texttt{desc} is the same caption-plus-full-source brief used in the TypedInk-internal ablation above.
This is the one acknowledged substitution --- the prompt text and execution harness are unmodified, and we did not rerun PaperBanana's Retriever, Stylist, or Critic agents, only the Visualizer in isolation, since those stages address figure style rather than the evidence-contract question this baseline targets.
Generation config matches PaperBanana's own Visualizer call site: temperature $0.2$, max output tokens $1800$, one shot (no retry, no repair loop --- PaperBanana's Visualizer does not repair failed code, so a failure is scored as a failure, not resubmitted).
Execution environment: Python 3.11.15, matplotlib 3.11.1, \texttt{Agg} backend, PaperBanana's unmodified \texttt{execute\_plot\_code\_worker} (\texttt{exec} in a fresh globals dict, no import allowlist).
Script: \texttt{evals/typedink/pb\_visualizer\_baseline.py}; raw per-view output: \texttt{evals/typedink/results/pb\_visualizer\_baseline.json}.

Result: $17/18$ views produce a chart ($1$ execution failure, \texttt{en\_bio\_exosome\_plot}/\texttt{struct}, from an unrestricted \texttt{import seaborn} that PaperBanana's prompt does not forbid).
No view recovers the full gold table (max per-view gold recall $0.5$; mean gold recall across inked views $0.044$).
Two of the $17$ inked views ink exactly one gold number string each despite the table being absent from the prompt: \texttt{en\_bio\_gelma\_plot}/\texttt{prose} inks \texttt{2.1} (one of two gold values, \texttt{3.4}/\texttt{2.1}), and \texttt{en\_mat\_modulus\_plot}/\texttt{struct} inks \texttt{112} (one of four gold values, \texttt{97}/\texttt{112}/\texttt{128}/\texttt{141}).
Neither view's prompt contains the table, so these are not table leakage; they read as plausible domain-typical guesses (fold-changes around 2--4$\times$, moduli around 100--150\,GPa) that happen to coincide with one gold value, which is consistent with our own definition of \emph{hallucinated} ink (a gold number string present in output whose source field was absent) rather than evidence the model saw the hidden table.

\bibliographystyle{unsrtnat}
\bibliography{references}

\end{document}